\documentclass[journal]{IEEEtran}

\usepackage{set space}
\usepackage{cite}
\usepackage{color}
\ifCLASSINFOpdf
   \usepackage[pdftex]{graphicx}
\else
  \usepackage[dvips]{graphicx}
   \DeclareGraphicsExtensions{.eps}
\fi
\usepackage{subfigure}

\usepackage{amsmath}
\usepackage{amsfonts,helvet}
\usepackage{amssymb}

\usepackage{algorithm}
\usepackage{algpseudocode}
\usepackage{array}
\usepackage{multirow}

\usepackage{hyperref}

\begin{document}
\title{Smart Small Cell with Hybrid Beamforming for 5G:\\ Theoretical Feasibility and Prototype Results}

% 교수님들 member 다시 확인할 것
\author{Jinyoung Jang, MinKeun Chung, Seung Chan Hwang, Yeon-Geun Lim, Hong-jib Yoon, TaeckKeun Oh, Byung-Wook Min, Yongshik Lee, Kwang Soon Kim, Chan-Byoung Chae,~and Dong Ku Kim

%\author{Jinyoung Jang,~\IEEEmembership{Student~Member,~IEEE},
%		  MinKeun Chung,~\IEEEmembership{Student~Member,~IEEE},
%		  \\Hae Gwang Hwang,~\IEEEmembership{Student~Member,~IEEE},
%		  Yeon-Geun Lim,~\IEEEmembership{Student~Member,~IEEE},
%		  \\Hong-jib Yoon,~\IEEEmembership{Student~Member,~IEEE},
%		  TaeckKeun Oh,~\IEEEmembership{Student~Member,~IEEE},
%		  \\Byoung-Wook Min,~\IEEEmembership{Member,~IEEE},
%		  Yongshik Lee,~\IEEEmembership{Senior~Member,~IEEE},
%		  \\Kwang Soon Kim,~\IEEEmembership{Senior~Member,~IEEE},
%		  Chan-Byoung Chae,~\IEEEmembership{Senior~Member,~IEEE},~and
%		  \\ Dong Ku Kim,~\IEEEmembership{Member,~IEEE}
%\IEEEauthorblockA{Yonsei University, Korea\\
%jiny.jang0@gmail.com,\{minkeun.chung, hwang819, yglim, recful, augustinooh, bmin, yongshik.lee, ks.kim, cbchae, dkkim\}@yonsei.ac.kr 
%}

% 홍집 이름 확인
\thanks{The authors are with Yonsei University, Korea (E-mail: \{dreaming\_hero, minkeun.chung, hwang819, yglim, recful, augustinooh, bmin, yongshik.lee, ks.kim, cbchae, dkkim\}@yonsei.ac.kr). C.-B. Chae and D. Kim are co-corresponding authors.}
\thanks{This research was partly supported by the MSIP (Ministry of Science, ICT and Future Planning), Korea, under the ``IT Consilience Creative Program" (IITP-
2015-R0346-15-1008) supervised by the IITP (Institute for Information \&
Communications Technology Promotion) and ICT R\&D program of MSIP/IITP (B0126-15-1012).}
\thanks{Full demo video is available at http://www.cbchae.org/}
%\thanks{The demo video is available at http://www.cbchae.org}
}

%\markboth{Transactions on Wireless Communications,~Vol.~, No.~, Month~Year}%
%{Shell \MakeLowercase{\textit{et al.}}: Massive MIMO Operation in Partially Centralized Cloud Radio Access Networks}


\maketitle
%%%%%%%%%%%%%%%%%%%%%%%%%%%%%%%%%%%%%%%%%%%%%%%%%%%%%
%%%%%%%%%%%%%%%%%%%%%%%%%%%%%%%%%%%%%%%%%%%%%%%%%%%%%
\begin{abstract}
In this article, we present a real-time three-dimensional (3D) hybrid beamforming for fifth generation (5G) wireless networks. One of the key concepts in 5G cellular systems is the small cell network, which settles the high mobile traffic demand and provides uniform user-experienced data rates. The overall capacity of the small cell network can be enhanced with the enabling technology of 3D hybrid beamforming. This study validates the feasibility of the 3D hybrid beamforming, mostly for link-level performances, through the implementation of a real-time testbed using a software-defined radio (SDR) platform and fabricated antenna array. Based on the measured data, we also investigate system-level performances to verify the gain of the proposed smart small cell system over long term evolution (LTE) systems by performing  system-level simulations based on a 3D ray-tracing tool.
\end{abstract}

\begin{IEEEkeywords}
Small cell, three-dimensional (3D) hybrid beamforming, fifth generation (5G) communications.
\end{IEEEkeywords}

% \fi
\IEEEpeerreviewmaketitle

%\newpage
%%%%%%%%%%%%%%%%%%%%%%%%%%%%%%%%%%%%%%%%%%%%%%%%%%%

%%%%%%%%%%%%%%%%%%%%%%%%%%%%%%%%%%%%%%%%%%%%%%%%%%%
\section{Introduction}
For the past several years, fifth generation (5G) wireless communication has been actively discussed both in academia and the industry as a means of providing various mobile convergence services.
Many 5G project groups have been established to propose service scenarios and target performance metrics. In the coming years, the amount of mobile data traffic is expected to increase tremendously~\cite{cisco2015}.
Given this fact, one important application that most 5G project groups are likely to support is high data rate communications~\cite{metis2015,5gforum2015}.
% Because of such a scenario, ITU recently agreed to define key performance indicators (KPIs) for 5G wireless communication including the user-experienced data rate and the data rate per area.

A key enabling technology to support a high data rate service scenario appears to be a small cell network \cite{metis2015,5gforum2015}.
When a network is in the interference-limited regime, it is known theoretically that spectral efficiency per area increases linearly with the number of small cells \cite{andrews2011}.
From a spatial domain viewpoint, however, there are huge differences can be found in the wireless channel characteristics of small cell networks (e.g., outdoor-to-indoor or indoor-to-indoor) and those of conventional macro cell networks.
Various channel environments may occur in a small cell network, since the space constraint for cell deployment can be greatly mitigated, thanks to the reduced deployment time and cost.
%For example, base stations (BSs) and users can be placed indoors or outdoors.
The channel conditions are affected dynamically by the relative locations among base stations (BSs) and users.
Furthermore, non-uniform cell deployment may cause crucial intercell interference (ICI). What is needed then is not only the densification of the network but also the small cell technologies that reflect the above issue so as to enhance the network performance and to fulfill 5G target requirements. 

In 5G cellular networks, a promising technology is one that exploits three-dimensional (3D) beam control.
In practical situations, BSs and users are distributed in 3D space, such as in the urban cell scenarios\cite{5gforum2015,Kim2014}.
As the elevation angle of the ray propagation becomes influential, the 3D beamforming can increase both the cell average throughput and the 5\%-tile user throughput\cite{Kim2014}.
A critical issue in addition to the 3D beamforming design is the performance evaluation method that reflects the 3D space accordingly. When it comes to 3D beamforming gain, more elaborate simulation results may be produced by generating BSs and users in both horizontal and vertical domains rather than considering only a two-dimensional (2D) distribution~\cite{Kim2014}.

  \begin{figure*}[!t]
    \centering
    \includegraphics[width = 6.8in]{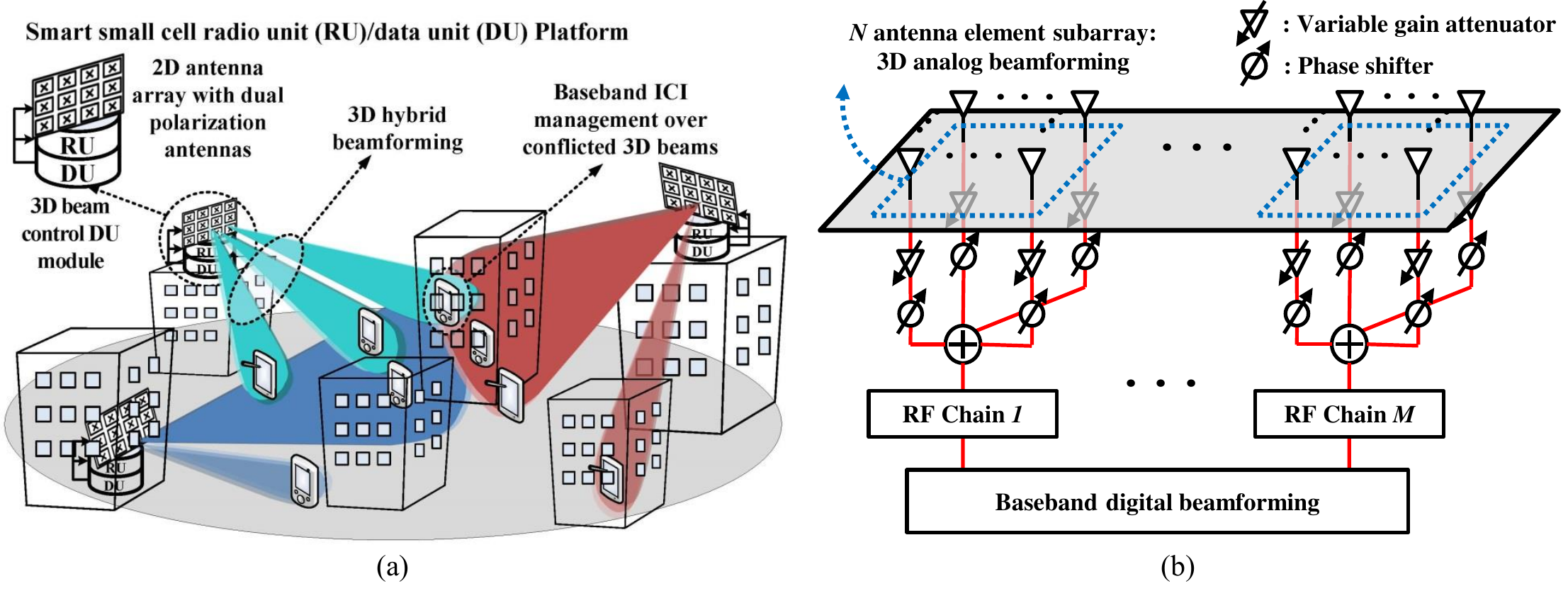}
    \caption{(a) Smart small cell concept, (b) 3D hybrid beamforming structure.}
    \label{fig:ssc_hbf}
    \end{figure*}

As a candidate for the small cell architecture of a 5G small cell network, we propose a smart small cell concept. In Fig.~\ref{fig:ssc_hbf}(a), the smart small cell system applies a mixture of 3D radio frequency (RF) analog beamforming and baseband digital beamforming--referred to as in this article as 3D hybrid beamforming. It provides user-perspective 3D beam directivity that is appropriate to the channel condition of each user. In order to exploit distinguishable beams at an elevation angle, the smart small cell system uses a 2D antenna array instead of a linear antenna array with a fixed radiation pattern in the vertical domain. When 3D beams among cells conflict, users can still experience a uniform data rate if additional baseband ICI management~\cite{Chae2012} is applied.

Focusing on a hybrid beamforming system, this smart small cell concept is proposed particularly for millimeter wave (mm-wave) communications. In the mm-wave communications, a beamforming with a number of antennas is used to overcome the pathloss of high frequency and to obtain a practical link budget. We further consider the hybrid analog/digital-processing strategy to reduce RF hardware costs and a computational complexity over the full digital beamforming that antenna elements are equipped with separate RF chains. The feasibility of the hybrid beamforming for mm-wave communications has been extensively evaluated by simulations and prototypes~\cite{xiaojing2010,roh2014}. 
In most cases, the beam weights are designed with the knowledge of instantaneous or statistical channel state information (CSI) at all antenna elements. This requires sufficient training before data transmission.

In this article, we discuss a novel 3D hybrid beamforming scheme designed without the advance knowledge of the CSI of all the antenna elements.
We examine the link- and system-level performances, through a real-time software-defined radio (SDR) testbed and 3D ray-tracing-based simulations, respectively, of the proposed smart small cell system and the conventional multiple input multiple output (MIMO) system. The results show that the proposed 3D hybrid beamforming system can enhance the area spectral efficiency of the 5G small cell network tremendously, even in the currently used frequency band of cellular systems (under 5~GHz).

%%%%%%%%%%%%%%%%%%%%%%%%%%%%%%%%%%%%%%%%%%%%%%%%%%
\section{3D Hybrid Beamforming Design} \label{sec:prop_hbf}

We consider a 3D hybrid beamforming system, shown in Fig.~\ref{fig:ssc_hbf}(b), that consists of $M$ RF chains and $M$ subarrays; a subarray is defined as a subset of antenna elements\cite{xiaojing2010}.
The total number of antenna elements is $MN$, such that each subarray has $N$ antenna elements and is connected to each RF chain (or baseband channel).
Note that the number of RF chains and that of antenna elements are the same in the conventional MIMO system whose antenna elements are connected to separate RF chains.
A 2D subarray is considered for both vertical and horizontal analog beamforming, i.e., directive 3D beamforming.
In the digital domain, multiple baseband channels can support multiple users, or spatial multiplexing. This can be done through highly directive beamforming by grouping several subarrays into a virtually large subarray.

A conventional hybrid beamforming design assumes perfect knowledge of the instantaneous CSI of all $MN$ antenna elements.
Such a design, however, is difficult to implement, as the training duration for sending pilot signals of such a hybrid beamforming system is at least $N$ times longer than that of the conventional MIMO system due to the lack of RF chains.
Furthermore, the system is incompatible with currently operating long term evolution (LTE), where it is difficult to estimate the CSI of all antenna elements before data transmission because the reference signal (RS) and the data are transmitted simultaneously over disjoint orthogonal frequency division multiplexing (OFDM) symbols.

In the proposed hybrid beamforming, the analog beamforming maximizes the $\it average$ signal-to-noise-ratio (SNR) of each subarray and the digital beamforming performs $M$-dimensional MIMO processing with $\it instantaneous$ CSI (comprising of analog beamweights and wireless channels). It is no trivial task to carry out the joint design of the analog beamforming and the digital beamforming with the absence of the CSI of all antenna elements. Hence we adopt a decoupled design of the analog beamforming and the digital beamforming. 
%This way we avoid having to measure the CSI of $MN$ antenna elements.
The proposed design for the analog beam weight targets the maximization of the average SNR of each subarray without the CSI of all antenna elements.
An iterative algorithm tracking the corresponding beam weight is applied. 
The proposed scheme only requires the information of the previously used beam weights and the corresponding $M$-dimensional beamformed baseband CSI.
For more details, see Section IV-B.
After the analog beam weight is computed, the $M$-dimensional digital beamforming, such as spatial multiplexing and diversity schemes, is applied based on the analog beamformed baseband CSI, which can be estimated with $M$ RF chains.
In this way, the proposed hybrid beamforming requires none of the CSI from the actual $MN$ antennas.

Also note that the physical layer of the LTE system can be directly applied without any modification into the proposed hybrid beamforming.
The number of antenna elements ($MN$) is larger than that of RF chains ($M$) in the hybrid beamforming system; nonetheless for the data modulation and demodulation in the baseband the proposed system does not require the information of the analog domain, such as the CSI of $MN$ antenna elements or the beamforming weight.
In other words, the $MN$-dimensional analog processing is transparent to the $M$-dimensional baseband and the increase of antenna elements does not vary any operation of the baseband processing in the physical layer.
% Moreover, even the beam weight computation operates independently of the data modulation and demodulation.
For more details, see Section IV-C.
Based on this architecture, also note that the antenna port concept of the LTE is compatible with the proposed hybrid beamforming system.
In the LTE downlink, several types of RS are provided and each RS pattern is transmitted from an antenna port at the BS.
Generally, each antenna port in the conventional LTE system consists of a single directive antenna element or multiple antenna elements using fixed beamforming.
According to the antenna port concept of the LTE, each antenna port in the proposed hybrid beamforming system may consist of a subset of subarrays using user-specific analog beamforming.
In either system, users can only estimate the composite channel of the antenna port regardless of the number of antenna elements making up the antenna port.

\section{Testbed Settings: System Specifications \& Hardware Architecture}\label{sec:testbed}
\begin{figure*}
    \centering
    \includegraphics[width = 7in]{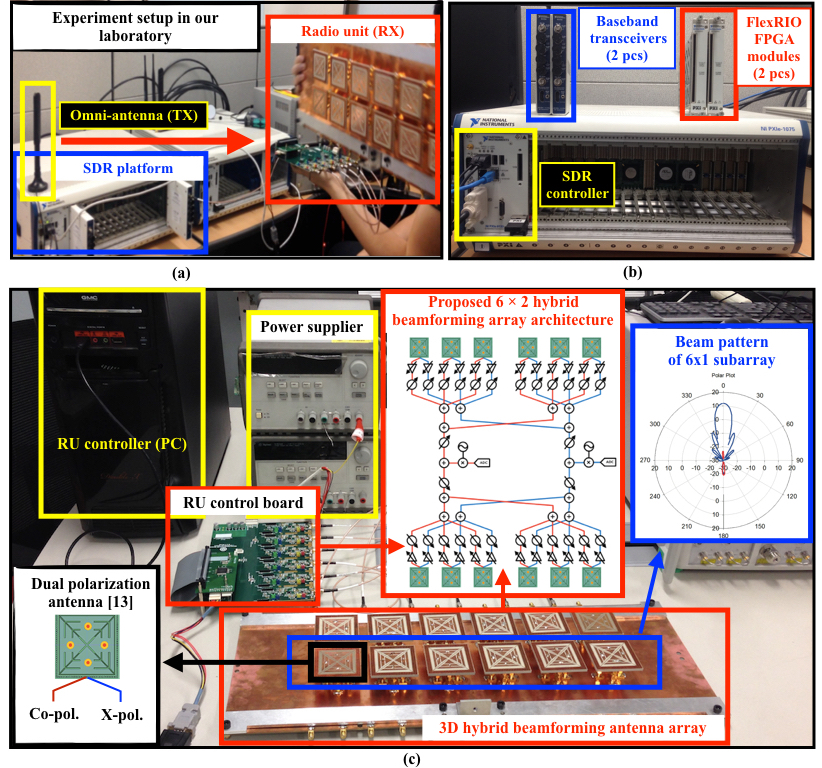}
    \caption{(a) Real-time hybrid beamforming testbed setup, (b) SDR platform, and (c) RU and supplementary hardware components.}
    \label{fig:testbed}
    \end{figure*}

The designed real-time hybrid beamforming testbed is based on the LTE standard \cite{sesia2009lte} with the following system specifications: a transmission bandwidth of 20~MHz, 30.72~MHz sampling rate, 15~kHz subcarrier spacing, 2048 fast Fourier transform (FFT) size, and variable 4/16/64 quadrature amplitude modulation (QAM). 
For this reason, we refer to the testbed as the hybrid beamforming-aided LTE system.
Figure~\ref{fig:testbed}(a) depicts the implemented prototype; a detailed explanation of the hardware components is given below.

\subsection{Software-Defined Radio Platform}
The SDR platform consists of the following hardware components, shown in 
Fig.~\ref{fig:testbed}(b).
\begin{itemize}
 \item {PXIe-8133}: Real-time controller equipped with a 1.73~GHz quad-core Intel Core i7-820 processor and 8~GB of dual-channel 1333~MHz DDR3 random access memory (RAM)\cite{ni8133}. 

 \item	{NI 5791R}: 100 MHz bandwidth baseband transceiver module equipped with dual 130~MSamples/s analog-to-digital converter (ADC) with 14-bit accuracy, and dual 130~MSamples/s digital-to-analog converter (DAC) with 16-bit accuracy\cite{ni5791}.

 \item	{PXIe-7965R}: Field-programmable gate array (FPGA) module equipped with a Virtex-5 SX95T FPGA optimized for digital signal processing, 512~MB of onboard RAM, and 16 direct memory access (DMA) channels for high-speed data streaming at more than 800~MB/s\cite{ni7965}.
\end{itemize}
In addition, NI modules sit in the NI PXIe-1075 chassis. The chassis is for data aggregation with both FPGA processors and a real-time controller for real-time signal processing.

\subsection{Radio Unit}\label{sec:radio_unit}
The radio unit (RU) is composed of a $6\times 2$ hybrid beamforming antenna array and an RU control board.
Figure~\ref{fig:testbed}(c) shows the RU and the supplementary components for measurement, such as the power suppliers and an RU controller (a PC) with a self-designed control program to run the processor embedded in the RU control board.

    \begin{figure*}[t]
    \centering
    \includegraphics[width = 7in]{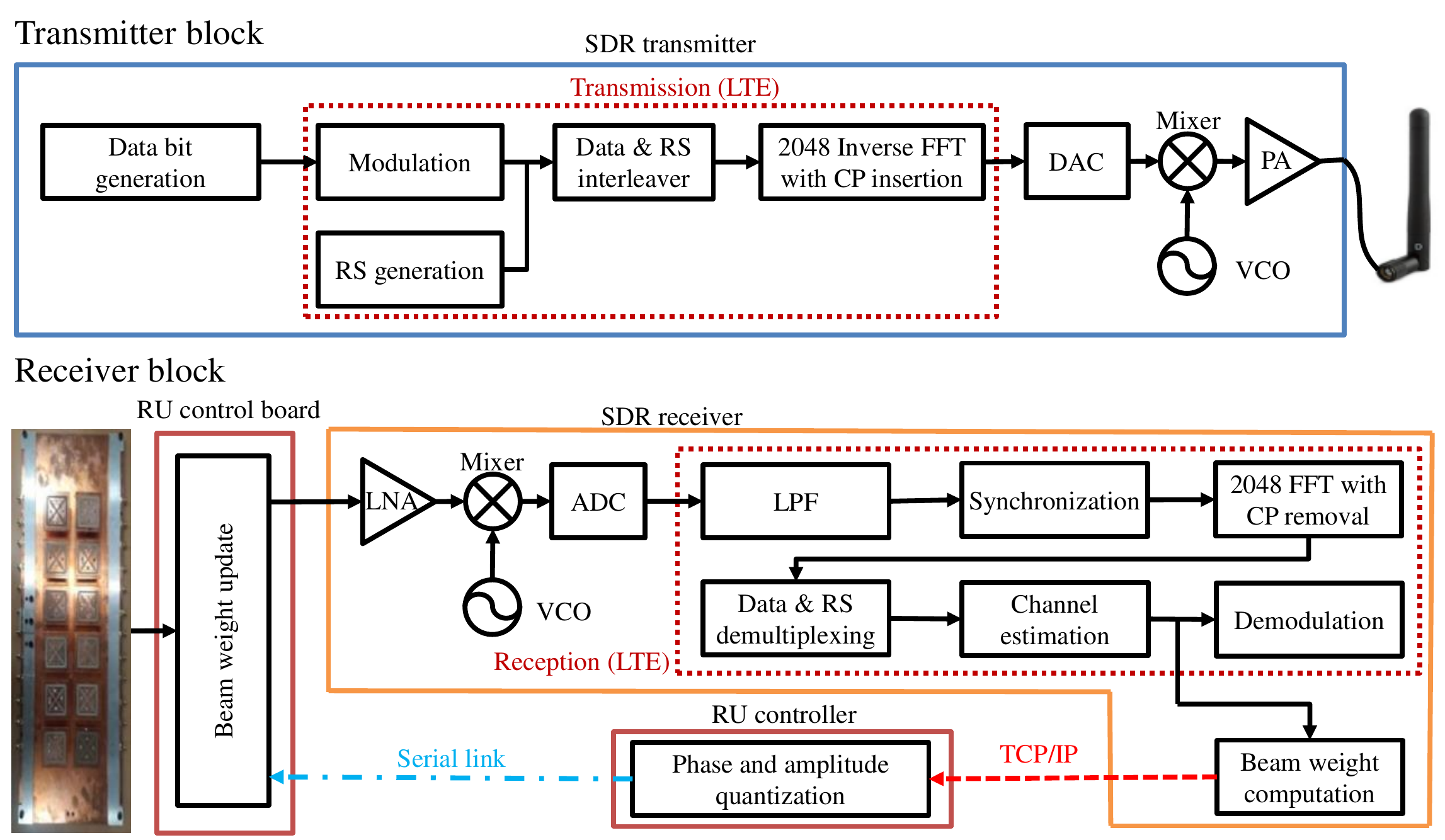}
    \caption{Block diagram of our real-time hybrid beamforming testbed.}
    \label{fig:block}
    \end{figure*}

The proposed 3D hybrid beamforming antenna array consists of two $6\times2$ subarrays based on the dual polarization antenna shown in~\cite{oh2014}.
The distance between neighbor dual polarization antennas is 0.6$\lambda$ where $\lambda$ is the wavelength.
The proposed design groups together $6 \times 2$ antenna elements with the same polarization on the RF control board as the subarray and connects them to a single baseband channel. This design reduces the size of the antenna array to half that based on antennas with a single polarization. 
Since, the 3D radiation patterns of Polarizations~1 and 2 of the dual polarization antenna in \cite{oh2014} are very similar, the beamforming patterns of the subarrays with different polarizations are nearly the same.
Most importantly, this antenna gives high cross-polarization discrimination (XPD) in all directions, where XPD is defined as the ratio of the co-polarized average received power to the cross-polarized average received power; it represents the separation of the radiation between Polarization 1 and Polarization~2. 
%High XPD is also shown in Fig.~\ref{fig:testbed}(c) in the beam pattern of the $6 \times 1$ antenna array that is a measured beam pattern of the $3 \times 1$ antenna array multiplied with a $2 \times 1$ array factor. 
High XPD is also shown in Fig.~\ref{fig:testbed}(c). The large difference between the main polarization (blue) and sub polarization (red) is seen in all directions implying that in the proposed design sufficiently suppresses the inter-subarray beam interference.

The RU control board is designed to control the phase and amplitude of signal input, for up to six antenna elements simultaneously.
The following hardware components are included:
\begin{itemize}
 \item ATMEGA2560: 8-bit processor for a 2~MHz channel
 
 \item MAPS-010144: Phase shifter (PS) of 4-bit quantization (quantization error: 11.25 degree) [6 pcs]
 
 \item HMC742LP5E: Variable gain attenuator (VGA) of 6-bit quantization (quantization error: 0.25~dB) [6~pcs]
 
 \item HMC715LP3 (RX RU): Low noise amplifier of 0.9~dB noise figure [6 pcs]
 
 \item MGA-22003 (TX RU): Power amplifier of 25~dBm output power [6 pcs].
\end{itemize}
In the RU, it takes approximately tens of $\mu s$ time delay at the processor to process the input beam weight and additional hundreds of $\mu s$ time delay at the six PSs and VGAs to update the beam weight (tens of $\mu s$ time delay per PS and VGA).
As a result, $1ms$ time delay, at most, is required to update the incoming beam weight into the hybrid beamforming antenna array. Such a delay is acceptable in the LTE since the uplink user scheduling is already determined at least $1ms$ earlier than the actual receiving time. Note that the system control delay is limited by the Transmission Control Protocol/Internet Protocol (TCP/IP), which is addressed in Section \ref{sec:prototype} and serial-link interface between the SDR receiver and the RU control board. This can be further optimized by implementing all the procedures in the FPGA chip. 

   \begin{figure*}[!t]
    \centering
    \includegraphics[width = 7in]{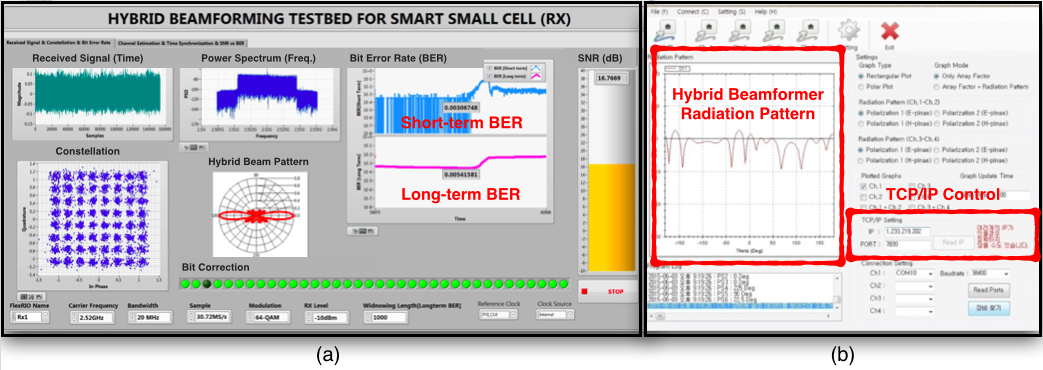}
    \caption{(a) A GUI of the SDR and (b) a GUI of the RU control program.}
    \label{fig:gui}
    \end{figure*}

\section{Real-time Hybrid Beamforming Testbed}\label{sec:prototype}
We now elaborate our design blocks for a real-time 3D hybrid beamforming testbed in a processing order.
We assume that there is one transmitter with an omni-antenna and one receiver with a hybrid beamforming antenna array.
The number of transmit and receive RF chains is one and the size of the subarray at the receiver is 6$\times$1. %\textcolor{red}{, i.e. $M=1$ and $N=6$ for the receiver}.
Figure~\ref{fig:block} illustrates the block diagram of the implemented testbed. 
We first briefly introduce the main procedure of the experiment and explain the detailed process of the main blocks in the following subsections.

\begin{enumerate}
\item
The SDR receiver randomly generates an initial beam weight and sends it to the RU controller through TCP/IP.
\item
The RU controller informs the RU control board of the beam weight, and the RU control board sets the beam weight.
\item
The SDR transmitter sends a signal in LTE Release~8 format with an omni-antenna and the SDR receiver equipped with a hybrid beamforming antenna array receives a beamformed signal through real radio channels.
\item
The SDR receiver performs synchronization, estimates the beamformed channel at the baseband, and decodes its desired signal.
\item
In parallel with Step 4, the SDR receiver computes a new beam weight that provides better average SNR by using the previously used beam weight and the currently beamformed channel.
\item
Proceed iteratively from Step 2 to Step 5.
\end{enumerate}

\subsection{Transmission and Reception}
The transmission and the reception are based on the LTE Release 8. An LTE frame has a duration of $10~ms$ and consists of 20 slots.
%    \begin{figure}
%    \centering
%    \includegraphics[width = 3.5in]{Figure/Fig5_v2}
%    \caption{Frame structure of our prototype.}
%    \label{fig:spec}
%    \end{figure}
Each slot has $0.5~ms$ of duration and 6 OFDM symbols with an extended cyclic prefix (CP). Although the test link in the SDR testbed is for the uplink, the LTE downlink frame structure is instead used for the simplicity of the testbed, where the RSs and the modulated data symbols are orthogonally mapped to the different subcarriers. The cell-specific RS of Antenna Port~0 in LTE is used. The sixth OFDM symbol of every first and eleventh slot of the frame includes a primary synchronization signal (PSS).
In LTE, the PSS uses Zadoff-Chu sequences for synchronization that has zero autocorrelation with its nonzero circular shift.
The length of the sequence is 63 in the frequency domain, and in the middle of the sequence is the DC-carrier with a null signal. For the reception where the PSS subcarriers need to be extracted from the received signal to calculate the correlation between the ideal sequence and the estimated PSS, we designed a low-pass filter (LPF) using Xilinx's finite impulse response (FIR) IP core. 
The designed LPF has a cut-off frequency of 1.4 MHz, a stop-band attenuation of 50 dB, and a pass-band ripple of 0.1 dB. 
%%%%
The correlation of PSS is calculated after the received signal samples pass through the LPF~\cite{FD2015}. %The index of the peak is the first sample index of the OFDM symbol~\cite{FD2015}.

For channel estimation, we implemented a linear interpolator using Xilinx's FIR IP core. 
%The RS subcarriers are extracted from the OFDM symbols after CP removal  and used to estimate the channel. 
%
% through a least square methodor the channel estimation in terms of a least square method. In order to estimate the channel coefficients of the data subcarriers, we implement
%
%For channel estimation, the RS subcarriers are extracted from the OFDM symbol after the CP removal.
%To the extracted RS subcarriers, a least square method is applied to estimate channel coefficients.
%Then to estimate the channel coefficients of the data subcarriers, we implement a linear interpolator using Xilinx's FIR IP core. 
The linear interpolator in each block estimates the channel coefficients of the data subcarriers as well as the RS subcarriers. 
The estimated channel coefficients are passed to the beam weight calculation block  (described in the next subsection) and to the demodulation block for the data symbols to be decoded.
For data decoding, a single-tap zero-forcing channel equalizer operates in each data subcarrier.

\subsection{Beam Weight Computation}\label{sec:beam_comp}
In this subsection, we explain the beam weight computation block.
For notational convenience, we assume that the number of baseband channels is one but can be readily generalized. 
%Also, note that we omit the subcarrier index for brevity.
As noted in Section \ref{sec:prop_hbf}, the proposed beam weight that maximizes the average SNR given by ${\mathbf{a}}^* = \arg\max\nolimits_{\mathbf{a},\|\mathbf{a}\|^2=1 } \mathbf{a}^H {\sf E}[\mathbf{h}\mathbf{h}^H]\mathbf{a} $ where $\mathbf{a}$ is the $N$-dimensional vector of candidate beam weight and $\mathbf{h}$ is the $N$-dimensional instantaneous CSI vector of antenna elements. 
It is obvious that the optimal beam weight is the dominant eigenvector of the spatial correlation matrix ${\sf E}[\mathbf{h}\mathbf{h}^H]$.
To obtain the spatial correlation matrix the instantaneous CSIs of the antenna elements must be measured and averaged out.

%. Therefore, it may not be a priori to either the transmitter or the receiver in the hybrid beamforming structure for the reasons given in Section~ \ref{sec:prop_hbf}. (XXX)

\begin{figure*}[t]
    \centering
    \includegraphics[width = 7in]{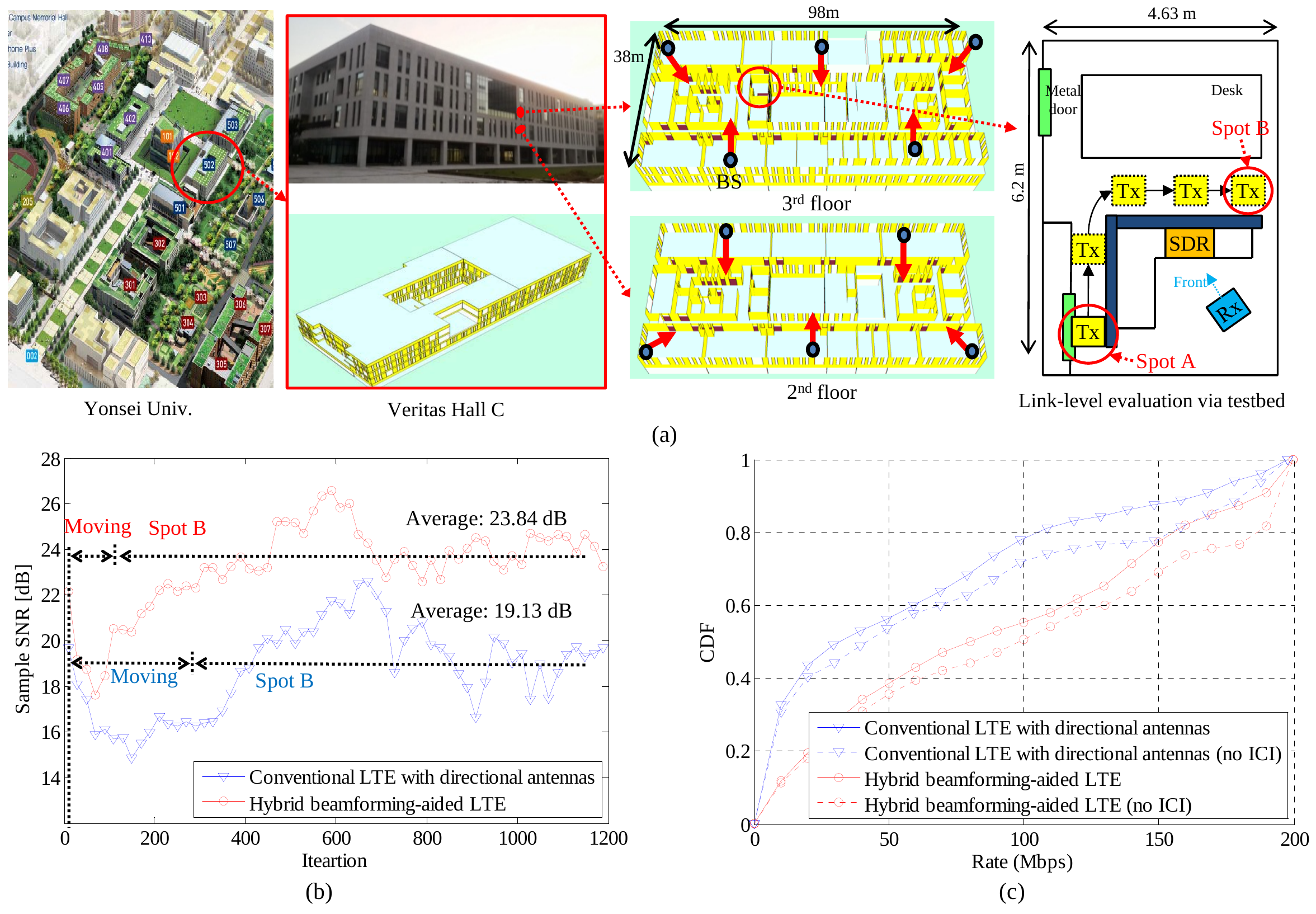}
    \vspace{-.05in}
    \caption{(a) A link- and system-level evaluation scenario, (b) a beam tracking performance of the link-level evaluation, and (c) a system-level evaluation result.}
    \label{fig:result}
    \vspace{-.05in}
\end{figure*}

The proposed algorithm is designed to iteratively track $\mathbf{a}^*$ without the instantaneous knowledge of either CSI for the antenna elements or the spatial correlation matrix.
Instead, we predict the beam weight with the history of the previously used beam weights and the corresponding received powers.
We denote the beam weight of the $i$th iteration to be $\mathbf{a}_i$, where an integer $i \ge 0$ and $\mathbf{a}_i$ is a unit vector. The beam weight computation procedure is as follows:
\begin{enumerate}
\item
Generate a slightly perturbed beam weight from the current beam weight and apply the perturbed weight for the signal reception. 

%Estimate the CSI of the baseband that is beamformed with $\mathbf{a}_i$ and compute the average received power of the beamformed channel.
\item 
Estimate the CSIs of each subcarrier in the baseband that is beamformed with the slightly perturbed beam weight during a subframe. 

%the baseband that is beamformed with the perturbation beam weights at $\mathbf{a}_i$ to all $N$ orthogonal complex dimensions and compute average received power of each perturbation beamformed channel.
%, $\mathbf{a}_{i,p}=\mathbf{a}_i + \beta\mathbf{p}_q$ for $q=1,\cdots,2N$, and compute average power of beamformed channel, which we denote $P_{i,q}$.
\item 
Compute the power difference among the perturbed beam weight and the previously used ones, and then find the unit vector $\mathbf{g}_i$ which is expected to give the maximum SNR growth at $\mathbf{a}_i$.
%Based on the history of measured power, $P_i$ and $\{P_{i,q}\}_{q=1}^{2N}$, compute the direction vector $\mathbf{g}_i$ for update which gives the maximum SNR growth rate at $\mathbf{a}_i$.
\item 
Compute a new beam weight $\mathbf{a}_{i+1}=\eta_i(\mathbf{a}_i+\alpha_i\mathbf{g}_i)$ where $\eta_i$ is chosen to normalize $\mathbf{a}_{i+1}$ and $\alpha_i$ is a properly chosen step size.
\end{enumerate}

Since the average SNR is a quadratic function with respect to $\mathbf{a}$, the maximization of which is a convex optimization.
Then, there exists a global optimal solution that can be found by an iterative algorithm such as the gradient descent method.
In the proposed algorithm, a gradient vector at $\mathbf{a}_i$ that is given by $\mathbf{g}_i$ is computed in Step 3 and the beam weight is updated to the gradient direction at $\mathbf{a}_i$ with proper scaling in Step 4.
The beam weight of the proposed algorithm converges near the optimal beam weight ($\mathbf{a}_{\infty} =\mathbf{a}^*$) and the step-response convergence requires at most a few-hundred subframes.

As also can be seen in Fig. \ref{fig:block}, the beam weight computation block changes operations in neither the LTE transmission nor the reception blocks. The proposed algorithm is compatible with the LTE system, and in our prototype design, the LTE system is simply extended to a 3D hybrid beamforming-aided system by including one additional processing block in the physical layer and the RU.

\subsection{Beam Weight Update Procedure in Radio Unit}\label{sec:beam_update}
After the beam weight computation is done, the following steps are carried out to update the beam weight in the RU.

\begin{enumerate}
\item The real-time controller sends the beam weight to the RU controller through TCP/IP.
\item The RU controller runs the RU control program that has the pre-measured array calibration table.
\item The RU control program reads the beam weight, quantizes the phase and amplitude of each coefficient according to the specifications of the PS and the VGA, and sends the bit sequences of the quantized phase and amplitude information to the processor of the RU control board. 
\item The processor of the RU control board commands PSs and VGAs to update the beam weights.
\end{enumerate}
Based on the overall time delay in processing these steps, we define the time period of the beam weight as follows.
In Step 1, the measured time delay of TCP/IP per link is $1~ms$, then the total time delay of TCP/IP links from the real-time controller to the RU controller via server becomes $\tau_1=2~ms$. In Steps 2 and 3, the measured processing time of the RU control program is, on average, $\tau_2 = 4~ms$. In Step~4, the time delay is $\tau_3=1~ms$ as mentioned in Section~\ref{sec:radio_unit}.
Overall, the total time delay is $\tau_1+\tau_2+\tau_3=7~ms$.
In a worst-case scenario, the beam-steering operation in RU might take place during an entire half frame since the duration of the half frame in LTE is $5~ms$. Hence, we set up the beam update in RU in such a way as to operate every 3 half frames ($15~ms$), abandoning the two half frames prior to data decoding so as to avoid the error in the beam transition time of the RU. Although the proposed algorithm is designed to update the beam weight at every subframe (i.e., 1 subframe per iteration), a longer period is used in the SDR testbed to avoid errors caused. Figure~\ref{fig:gui} illustrates graphical user interfaces (GUIs) of the testbed controllers (SDR and RU control board) and a part of source code for the FPGA design.\footnote{Full demo video is available at http://www.cbchae.org/. }

\section{Evaluation Results}

In this section, we investigate the link- and system-level performances of the small cell networks of the conventional LTE system and the proposed hybrid beamforming-aided LTE system. 
We assume that the BSs in the conventional LTE system use $M$-dimensional antennas in \cite{oh2014} and $M$ RF chains.
As the BSs use a single directional antenna per RF chain, the performance of the conventional LTE with a directional antenna is the same as the performance of spatial multiplexing without hybrid beamforming.
The simulation parameters refer to the LTE specifications, which mirror the prototype settings in Section~\ref{sec:testbed}.

Figure~\ref{fig:result}(a) illustrates our evaluation scenarios.
For link-level evaluations, we tested beam tracking performances of the hybrid beamforming-aided LTE system using the real-time testbed described in Section IV where the number of RF chain is one and the size of the subarray is 6$\times$1, i.e., $M=1$ and $N=6$.
An indoor-to-indoor channel in a rich scattering environment was considered.
We measured the received SNR at the receiver in Spot A and Spot~B--locations of equal height.
The proposed hybrid beamforming-aided LTE system showed approximately 4$\sim$4.7~dB higher received SNR (see Table~\ref{table:lle}), on average, than the conventional LTE system. Figure~5(b) shows the measured sample SNR of a trial test. The transmitter with an omni-antenna transmitted to the hybrid beamforming receiver with the RU; the transmitter was moved from Spot A to Spot B. The proposed scheme tracked the beam well even when the angular variation of the channel changed rapidly in the trial test. One might argue that this SNR gain might not enough to meet the 5G requirements; the purpose of this SDR testbed was to show that the proposed 3D hybrid beamforming performs pretty well in real environments.  

\begin{table}[t]
	\caption{Performance comparison results between the SDR testbed and the 3D ray-tracing (WiSE) simulation.}
	\centering
	\label{table:lle}
	\vspace{-0.1in}
   \includegraphics[width = 3.3in]{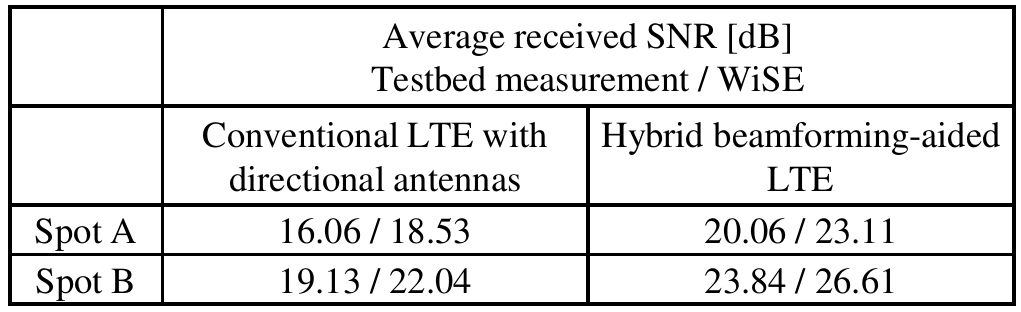}
 	\vspace{-0.2in}
\end{table}

%% System Level Simulation result
%So far, we have evaluated the smart small cell system through the real-time SDR testbed. 
To evaluate the system level performance gain, we developed Wireless System Engineering (WiSE)~\cite{val1998} (a 3D ray-tracing tool developed by Bell Labs) based system level simulator.  As described in Table~\ref{table:lle}, the trend of the measured SNR of the link-level evaluation is similar to the that of the WiSE simulation. Note that the SNR differences between the testbed result and WiSE simulation are plausible since the WiSE simulations omitted furniture and small scatters. Here we evaluated the downlink system throughput performance of the conventional LTE system and the 3D hybrid beamforming-aided LTE system in Veritas Hall C of Yonsei University in Korea. Five small cell BSs were deployed on the second and third floors respectively according to the locations shown in Fig.~\ref{fig:result}(a). 
The downlink rate evaluation per mobile station (MS) was then carried out.
The MSs were uniformly distributed on every floor of the buildings and they were associated with the cell that provided the strongest SNR.
For the proposed hybrid beamforming-aided LTE system, we assumed that the uplink beam weight is reciprocal to the downlink beam weight, and the BS used the saturated beam weight.
For both systems, the small cell BSs had two transmit RF chains and the MSs had two receive RF chains. 
For the proposed hybrid beamforming-aided LTE, the number of RF chain was two and the size of the subarray was 6$\times$2 as described in Fig. \ref{fig:testbed}(c), i.e., $M=2$ and $N=12$.
Each cell randomly scheduled one MS in the coverage and performed single-user MIMO with spatial multiplexing (LTE Transmission Mode~4).

Figure~\ref{fig:result}(c) illustrates the cumulative density function of the ergodic achievable rate of the MSs when the transmit power of the BSs was 23~dBm. 
The system parameters follow the LTE specifications where the system bandwidth was 20~MHz; thus the maximum data rate is 200~Mbps~\cite{sesia2009lte}.
We also depicted the achievable rates of both systems while assuming no ICI.
%The high directive beam with a low side lobe level decreased the inter-beam interference powers even as it increased the desired power. 
The average rates of the conventional LTE system and the 3D hybrid beamforming-aided LTE system were 56.97~Mbps and 89.04~Mbps respectively.
This gave rise to an enhanced rate performance, approximately 56\% on average, for the entire service area.

\section{Conclusion}
This article has proposed a smart small cell concept to play a key role in supporting 5G networks, in which a user-specific 3D hybrid beamforming. In this article, we first validated the feasibility of the proposed 3D hybrid beamforming by implementing a real-time SDR testbed. Based on the measured data through the SDR testbed, we also performed 3D ray-tracing-based system-level simulations to investigate the system-level potential gain of the proposed smart small cell system. 
We expect our prototype design to provide worthwhile insights into developing the most viable solution for future wireless communication systems with an in-depth consideration of practical implementation.

%As a major role to support 5G cellular systems, the smart small cell system applies the 3D hybrid beamforming to increase overall system throughput and the intercell interference management scheme over conflicted beams to provide a uniform user experienced data rate.
%In this article, we validate the feasibility of the 3D hybrid beamforming by implementing a real-time testbed using SDR platform, and perform the system level simulation using 3D ray tracing tool to investigate the potential gain of the proposed system.

%\section*{Acknowledgment}


\ifCLASSOPTIONcaptionsoff
  \newpage
\fi




%\bibliographystyle{IEEEtran}

\renewcommand{\baselinestretch}{1.0}
\bibliographystyle{IEEEtran}
\bibliography{reference_CommMag14} % file name

%
%
%
%% biography section
%% 
%% If you have an EPS/PDF photo (graphicx package needed) extra braces are
%% needed around the contents of the optional argument to biography to prevent
%% the LaTeX parser from getting confused when it sees the complicated
%% \includegraphics command within an optional argument. (You could create
%% your own custom macro containing the \includegraphics command to make things
%% simpler here.)
%%\begin{IEEEbiography}[{\includegraphics[width=1in,height=1.25in,clip,keepaspectratio]{mshell}}]{Michael Shell}
%% or if you just want to reserve a space for a photo:
%
%%\epsfysize=3.2cm

%\vspace{-.5in}
\begin{IEEEbiographynophoto}{Jinyoung Jang} [S'10] received his B.S. degree from the School of Electrical and Electronic Engineering at Yonsei University, Korea, in 2009. He is now working toward his Ph.D. degree at the same university. His research interests include the design and implementation of architectures for next-generation wireless communication systems.
\end{IEEEbiographynophoto}

%\vspace{-.5in}
\begin{IEEEbiographynophoto}{MinKeun Chung} [S'11] received his B.S. degree from the School of Electrical and Electronic Engineering at Yonsei University, Korea, in 2010. He is now working toward his Ph.D. degree under the joint supervision of Prof. D. K. Kim and Prof. C.-B. Chae. He did his graduate internship in the Advanced Wireless Research Team at National Instruments, Austin, Texas in 2013 and 2015. His research interests include the design and implementation of architectures for next-generation wireless communication systems.
\end{IEEEbiographynophoto}

%\vspace{-.5in}
\begin{IEEEbiographynophoto}
{Seung Chan Hwang} [S'15] was born in Ulsan, Korea, on August 19, 1982. He received the B.S. and M.S. degrees from Yonsei University, Seoul, Korea, in 2005. and 2007. He is currently working toward the Ph.D. degree in the Department of Electrical and Electronic Engineering, Yonsei University. His research interests include multiuser MIMO and mmWave communications.
\end{IEEEbiographynophoto}

%\vspace{-.5in}
\begin{IEEEbiographynophoto}{Yeon-Geun Lim} [S'12] received the B.S. degree in
Information and Communications Engineering from Sungkyunkwan University, Korea, in 2011. He is currently working toward the Ph.D. degree in the School of Integrated Technology, Yonsei University. His research interest includes massive MIMO and interference management techniques for smart small-cell networks.
\end{IEEEbiographynophoto}

%\vspace{-.5in}
\begin{IEEEbiographynophoto}{Hong-jib Yoon} [S'14] received his B.S. degree from the School of Electrical and Electronic Engineering at Yonsei University, Korea, in 2013. He is now working toward his Ph.D. degree at the same university. His research interests include the design and implementation of passive devices and phase array for wireless communication systems.
\end{IEEEbiographynophoto}

%\vspace{-.5in}
\begin{IEEEbiographynophoto}{TaeckKeun Oh} was born in Seoul, Korea. He received the B.S degree in Electronic Engineering from Inha University, Incheon, Korea, in 2010, the M.S degree in Electrical and Electronic Engineering from Yonsei University, Seoul, Korea, in 2015. Since 2015, he has been with LIG Nex1, Seong-Nam, Korea, as a research engineer. His work is focused on satellite communication system. His research interests include active phased array system, multi-polarization antenna, active/passive components in the field of RF and microwave, and embedded systems.
\end{IEEEbiographynophoto}

%\vspace{-.5in}
\begin{IEEEbiographynophoto}{Byung-Wook Min} [M'08] is an assistant professor with the Department of Electrical and Electronic Engineering, Yonsei University, Seoul, Korea. He received the B.S. degree from Seoul National University, Seoul, Korea, in 2002, and the M.S. and Ph.D. degrees from The University of Michigan at Ann Arbor, in 2004 and 2007, respectively. In 2006-2007, he was a visiting scholar with the University of California at San Diego, La Jolla. In 2008-2010. he was a senior engineer at Qualcomm Inc., Santa Clara, CA and Austin, TX. His research interests include Si RFIC and communication systems for microwave and millimeter-wave applications. He was a recipient of the Samsung Scholarship in 2002-2007.
\end{IEEEbiographynophoto}

%\vspace{-.5in}
\begin{IEEEbiographynophoto} %(S’00-M’04-SM’12)
{Yongshik Lee} [SM'12] was born in Seoul, Korea. He received the B.S. degree from Yonsei University, Seoul, Korea, in 1998, and the M.S. and Ph.D. degrees in electrical engineering from The University of Michigan at Ann Arbor, in 2001 and 2004, respectively. In 2004, he was a Postdoctoral Research Associate with Purdue University, West Lafayette, IN. From 2004 to 2005, he was with EMAG Technologies Inc., Ann Arbor, MI, as a Research Engineer. In September 2005, he joined Yonsei University, Seoul, Korea, and is currently a Professor. His current research interests include passive and active circuitry for microwave and millimeter-wave applications, and electromagnetic metamaterials.
\end{IEEEbiographynophoto}

%\vspace{-.5in}
\begin{IEEEbiographynophoto}{Kwang Soon Kim} [SM'04] % (S'95, M'99, SM'04) 
 received the B.S. (summa cum laude), M.S.E., and Ph.D. degrees in electrical engineering from KAIST, Korea. From March 1999 to March 2000, he was with the Dept. of Electrical and Computer Engineering, University of California at San Diego as a Postdoctoral Researcher. From April 2000 to February 2004, he was with the Electronics and Telecommunication Research Institute, Daejeon, Korea as a Senior Member of Research Staff. Since March 2004, he has been with the Dept. of Electrical and Electronic Engineering, Yonsei University, Seoul, Korea, is currently a Professor.
Prof. Kim is a Senior Member of the IEEE, served as an Editor of the Journal of the Korean Institute of Communications and Information Sciences (KICS) from 2006-2012, as the Editor-in-Chief of the journal of KICS since 2013, as an Editor of the Journal of Comm. and Networks (JCN) since 2008, as an Editor of the IEEE Trans. on Wireless Comm. 2009-2014. He was a recipient of the Jack Neubauer Memorial Award (Best system paper award, IEEE Trans. Vehicular Tech.) in 2008, and LG R\&D Award: Industry-Academic Cooperation Prize, LG Electronics, 2013. %His research interests are in signal processing, communication theory, information theory, and stochastic geometry applied to wireless heterogeneous cellular networks, wireless local area networks, wireless D2D networks and wireless ad doc networks.
\end{IEEEbiographynophoto}
\vspace{-.5in}
\begin{IEEEbiographynophoto}{Chan-Byoung Chae} [SM'12] is an associate professor in the School of Integrated Technology, Yonsei University. Before joining Yonsei University, he was with Bell Labs, Alcatel-Lucent, Murray Hill, New Jersey, as a member of technical staff, and Harvard University, Cambridge, Massachusetts, as a postdoctoral research fellow. He received his Ph.D. degree in electrical and computer engineering from The University of Texas at Austin in 2008. He was the recipient/co-recipient of the Best Young Professor Award from the College of Engineering, Yonsei (2015), the IEEE INFOCOM Best Demo Award (2015), the IEIE/IEEE Joint Award for Young IT Engineer of the Year (2014), the KICS Haedong Young Scholar Award (2013), the {\it IEEE Signal Processing Magazine} Best Paper Award (2013), the IEEE ComSoc AP Outstanding Young Researcher Award (2012), the IEEE Dan. E. Noble Fellowship Award (2008), and two Gold Prizes (1st) in the 14th/19th Humantech Paper Contest. He currently serves as an Editor for {\it IEEE Trans. Wireless Comm.}, the {\it IEEE/KICS Jour.  Comm. Networks}, and {\it IEEE Trans. Molecular, Biological, and Multi-scale Comm}.
\end{IEEEbiographynophoto}

%\vspace{-.5in}
\begin{IEEEbiographynophoto}{Dong Ku Kim} [SM'15] received his B.S. from Korea Aerospace University in 1983, and his M.S. and Ph.D. from the University of Southern California, Los Angeles, in 1985 and 1992, respectively. He worked on CDMA systems in the cellular infrastructure group of Motorola at Fort Worth, Texas, in 1992. He has been a professor in the School of Electrical and Electronic Engineering, Yonsei University, since 1994, and was the principal investigator professor of the Qualcomm Yonsei Joint Research Program from 1999 to 2010. Currently, he is a vice president for academic research affairs of the KICS, and a vice chair of 5G Forum. He received the Minister Award for the Distinguished Service for ICT R\&D from the MISP in 2013, and the Award of Excellence in leadership of 100 Leading Core Technologies for Korea 2020 from the National Academy of Engineering of Korea.
\end{IEEEbiographynophoto}

%
%
%


%\vfill


\end{document}